\documentstyle[preprint,aps]{revtex}
\newcommand{\nin}{\noindent}
\newcommand{\be}{\begin{equation}}
\newcommand{\ee}{\end{equation}}
\newcommand{\ber}{\begin{eqnarray}}
\newcommand{\eer}{\end{eqnarray}}

\begin{document}
 
\title{Bragg resonances
        for tunneling between edges of a 2D Quantum Hall system}
\author{M. B. Hastings and  L. S. Levitov}
\address{Physics Department, Massachusetts Institute of Technology}
\maketitle
 
\begin{abstract}
A theory is presented for tunneling between compressible regions
on the sides of a narrow incompressible Quantum Hall strip.
Assuming that electron interactions lead to formation of a
Wigner crystal on the edges of the compressible regions, we
consider the situation when the non-conservation of electron
momentum required for transport is provided, in the absence of
disorder, by umklapp scattering on the crystal. The momentum
given to the crystal is quantized due to the Bragg condition,
which leads to resonances in tunneling conductivity as a
function of the incompressible strip width, similar to those
reported recently by N.~Zhitenev, M.~Brodsky, R.~Ashoori, and M.~Melloch.
   \end{abstract}
 
The structure of the edge of a Quantum Hall system is
central for understanding the QH
physics\cite{edge-current-review}. Theories of the QH edge
predict that the edge states form an interesting strongly
interacting one-dimensional system. Different aspects of the
edge dynamics are described by the chiral Luttinger liquid
theories\cite{chiral-Luttinger}, and by theories taking into
account the long-range character of electron
interaction\cite{Mitya}, exchange\cite{exchange} and correlation
effects\cite{edge-reconstruction}. One expects that the QH edge
has a rich structure. To probe it, one can study electric
transport across the edge, e.g., by tunneling into the
edge\cite{tunneling}, or by measuring an ac response. The latter
was realized in a recent ac capacitance measurement\cite{Ray},
in which a capacitance probe attached above the 2D gas was used
to accumulate charge under the probe and to form a sharp
gradient of electron density near the probe edge. When the
densities in the 2D gas and under the probe are greater and less
than an integer, correspondingly, an incompressible strip is
formed near the probe edge. Since electric field near the edge
is strong, it was possible to create a very narrow
incompressible strip of width estimated as 8--10 magnetic
lengths.
 
Measuring the frequency dependent ac capacitance gives tunneling
conductivity across the strip. As a function of the potential
$V_g$ on the probe, the conductivity displays resonances,
uniformly spaced in $V_g$\cite{Ray}. The resonances were studied
at different magnetic fields and temperatures. The resonances
shift as the field changes, however, the shift is much weaker
than corresponding change of filling fraction, i.e., the
resonances are more sensitive to $V_g$ than to the field. It was
found that the resonances broaden with increasing temperature,
and that the peak values of conductivity increase roughly as
$T^{\alpha}$, with $\alpha\ge 1$.
 
It was demonstrated that the peaks are not caused by resonance
tunneling through impurities: the number of peaks is about the
same for the probes of different sizes, and the conductivity at
the peaks scales as the probe size. The high degree of the peaks
ordering, as well as the temperature dependence, inconsistent
with resonance tunneling, suggests looking for a different
mechanism. Since the strip is narrow, it may be possible to
tunnel across it without scattering off of impurities. Momentum
conservation then requires that the tunneling
electron give up momentum to
other electrons. If the other electrons (or holes in the region
of less than integer filling factor) form a Wigner
crystal\cite{FertigI,Nazarov}, momentum can only be given up at
multiples of the inverse lattice spacing. This effect would lead
to Bragg resonances in tunneling conductivity when these
multiples coincide with the momentum required to tunnel across
the strip.
 
In all theories of the QH edge predicting structure of a few
magnetic lengths wide\cite{Mitya,exchange,edge-reconstruction},
there is room for a 1D Wigner crystal. For example, the theory
of edge reconstruction\cite{edge-reconstruction} leads to an
edge composed of Luttinger liquids, one chiral, and several
non-chiral, each of the latter representing nothing but a 1D
crystal. One expects no long-range order in a 1D system, so it
is equally correct to view a 1D crystal as a correlated fluid.
The correlation length, however, varying as the inverse of
temperature, becomes large at low temperatures. This leads to
finite widths of the peaks in the structure factor, scaling with
temperature.
 
We will discuss a Wigner crystal with
varying density, formed near the incompressible edge, and
study tunneling into such a crystal. For the edge structure we use
the long-range screening picture\cite{Mitya}, however,
since only one row of the crystal will be important, our predictions will
hold for any reconstructed edge.
We consider Bragg resonances
in tunneling conductivity, and calculate the dynamical
Debye-Waller factor due to the crystal zero-point and thermal
motion, giving the temperature dependence of the peaks. The peak
intensities, thermal broadening, and shifts caused by varying
temperature, are found to be consistent with the experimental
observations\cite{Ray}.
 
\nin \underline{\it Wigner crystal at the edge:} \hskip2mm
In magnetic field, electrons form a crystal at
sufficiently low densities and temperatures, where potential
energy overcomes kinetic energy\cite{Fukuyama}. The critical
density $\nu$ was estimated to be slightly below
$1/5$\cite{critical-density}. At higher densities a QH fluid
will form.
 
On the sides of an incompressible strip, formed by one or
several fully filled Landau levels with integer density, the
electron density is non-integer and varies smoothly on a scale,
set by long-range electric forces, of many magnetic
lengths\cite{Mitya}. Thus the density of excess electrons and
holes right near the strip must be quite small. Within the two
small density regions there are several crystalline rows of
electrons and holes, respectively (see Fig.~\ref{fig1}.b). The
number of the rows is determined by the external electric field
$E$. In the continuum limit\cite{Mitya}, the electron density
measured from the integer within the strip goes to zero at some
point with a square-root singularity. The discrete crystal will
have its front row located near that point. The rows separation
can be determined by balancing forces, $eE$ from confining
potential against the force $e^2/\epsilon r^2$ from neighboring
particles, which, up to a factor $\sim1$, gives the spacing in
the front row: $a=\sqrt{e^2/\epsilon E}$. In the
experiment\cite{Ray}, the e- and h-crystals both have small
density at the strip edge, and increase away from it. We
estimate that $a$ is of the order of 2-3 magnetic lengths for
the front row of each crystal. In the continuum
limit\cite{Mitya}, the particle spacing goes as the $-{1\over4}$
power of the distance from the edge. Similarly, in numerical
simulations\cite{FertigI}, the particle spacing varies very
slowly with the distance from the edge. However, the estimated
density of the front row is close to the Wigner crystal critical
density\cite{critical-density}, and so we expect that beyond a
few crystal rows a QH fluid is formed (Fig.~\ref{fig1}.b shows
e- and h-rows obtained by a simulation similar to
\cite{FertigII}).
 
At constant density, a Wigner crystal forms a perfect triangular
lattice. In the presence of the field forming the strip, the
density of the crystal will increase along the field, however,
because of the long-range interaction, the crystal forms
well-ordered rows, even at the edge\cite{FertigI}. The largest
simulations of a crystal in an external field that we know of
were done by Pieranski using $1/r^3$-interacting
dipoles\cite{Pieranski} (see Fig.~\ref{fig1}.a for a picture of
the resulting crystal). These experiments also revealed the
remarkable idea of the conformal crystal\cite{Pieranski}. Since
the electron interaction $1/r$ is longer-range than $1/r^3$, the
front rows in the $1/r$ crystal are ordered even better than in the
$1/r^3$ crystal (see Fig.~2 in \cite{FertigII}).
 
We expect that both electrons and holes will be surrounded by
spin textures, topological excitations of the local
spin\cite{skyrmions}. We will ignore any effect this has on the
crystal, as well as any question of how the spin is transferred
across the strip. It may be that spin is transferred
independently of charge, by spin waves.
 
\nin \underline{\it Bragg reflection}\hskip2mm
Since the tunneling amplitude falls exponentially with distance,
the main contribution to the tunneling rate arises from
transitions between the front e- and h-rows
(Fig.~\ref{fig1}.b). Therefore, we will focus attention on only one
row on each side of the strip. We consider a strip of width $w$
going in the $x-$direction, with e- and h-crystal row spacing
$a$. In Landau gauge, electron states can be written as
  \be\label{w-function}
\psi(x,y) = e^{i p x} f_n(y + pc\hbar /eB)
  \ee
where $f_n$ is the harmonic oscillator $n-$th wave function, $p$
labels the $x$-momentum.
 
The tunneling amplitude is given by the overlap of states
(\ref{w-function}) on opposite sides of the strip. In the
absence of momentum non-conserving scattering, the amplitude
will vanish because shifting $y$ by $w$ across the strip is
equivalent to a change of $p$, but the states with different
$p$'s are orthogonal. In the presence of a crystal, electron
momentum may change by a multiple of the Bragg vector,
$\delta p=2\pi n/a$. (For simplicity, let us assume that the
crystal period is the same on both sides of the strip.) On the
other hand, since according to (\ref{w-function}) the change in
momentum is coupled to a $y-$shift, the overlap of the states
will be maximal at $\delta p c\hbar /eB=w$. The two conditions
put together give that tunneling must be maximal at
  \be\label{peaks}
w_n=\ {\Phi_0\over aB}\ n\ ,
  \ee
where $n$ corresponds to the Bragg resonance order, and
$\Phi_0=hc/e$ is the flux quantum.
 
 
From (\ref{peaks}), the peak positions are dependent upon the
strip width $w$, and thus are very sensitive to the gate voltage
$V_g$. Increasing $V_g$ leads to squeezing of the strip:
roughly, $w$ is inversely proportional to $V_g$. For this
reason, we expect that the dependence on $V_g$ of the $B-$field
needed to observe a given resonance is stronger than the
dependence on $V_g$ of the $B-$field needed to maintain a given
filling fraction under the gate. This is seen in the experiment.
 
To estimate the order of a resonance one takes the product
$wa/l_H^2$, where $l_H$ is magnetic length. In the experiment,
$l_H\simeq 10$nm, $w$ is about $5-8\ l_H$, and $a$ is estimated
above to be $2-3\ l_H$. This gives the order of resonance
between 10 and 25, consistent with the experiment.
 
\nin \underline{Going from a 2D crystal to single row:}\hskip2mm
To estimate the response of the crystal to injecting an
electron, we assume that all relaxation occurs within one row,
and treat it as a one-dimensional system. The other electrons
will be taken into account only to the extent that they produce
a confining potential $V(y)$ for that row. This approximation
underestimates the crystal stiffness, enhanced by the presence
of other rows, and thus overestimates the peak widths. However,
we believe the discrepancy is not large, because for a crystal
with long-range interaction the shear modulus is much smaller
than the compressibility, and also the particles in the front
row, being more widely spaced, are more loose than in other
rows. Thus, charge spreads most easily within the first row.
 
With this, one writes the Hamiltonian for one row as
$$
\widehat{\cal H}=\sum\limits_n
{(p_{n,x}-{e\over c}By_n)^2 \over 2m}+
{p^2_{n,y}\over 2m} + V(y_n)
+{1\over2}\sum\limits_{m,n} U(r_n-r_m)\ ,
 $$
where $r_n=x_n\hat i +y_n\hat j$ are electron positions, and
the interaction $U(r)= e^2/\epsilon|r|- e^2/\epsilon
(r^2+d^2)^{1/2}$ accounts for the screening by the gate.
 
In a magnetic field, kinetic energy is very small
compared to potential energy, and so the fluctuations about mean
positions in the crystal will be small. We assume that different
Landau levels are not mixed by the interaction $U$, and then
treat the displacements $x_n$ and $y_n$ as conjugate variables
with the commutation relation $[x_n,y_n]=i\Phi_0/B$. Expanding
the Hamiltonian, and keeping only quadratic terms, one obtains a
harmonic chain problem. Diagonalizing the Hamiltonian, one gets
elementary excitations. Their
spectrum at small $qa\ll 1$ is acoustic: $\omega(q)=c|q|$,
where $c=a\sqrt{\kappa/m_*}$. (Here $\kappa=\sum_n n^2
U''_{r=an}$, $m_*=m(1+m\omega_c^2/V''_{r=0})$.) At small $q$,
the displacements are along the row.
 
\nin \underline{Evaluating the tunneling rate:}\hskip2mm
The tunneling is described by transferring an electron across
the strip, from the e-row to the h-row. Injecting an electron
into the row is accompanied by motion of other electrons that
give space to the new electron. This makes the tunneling a
collective process, and requires considering relaxation in the
row. Since at high magnetic field the effective mass of
electrons in the row is large, $m_*\gg m$, the relaxation will
be slow. This simplifies analytic treatment, making it possible
to consider only the effect of acoustic modes, $qa\ll 1$. In
order to evaluate the tunneling rate we assume that the transfer
of an electron across the strip is a much faster process than
charge relaxation within the row, and that the characteristic
time scales are much larger than $a/c$. Such an assumption is
self-consistent at low $T<\hbar\omega_c$, which is within
experimental range (see similar discussion
in~\cite{string-breaking}).
 
We evaluate the tunneling rate semiclassically, by taking the
saddle-point of the action for the dynamics in imaginary time
corresponding to injection and removal of an electron in the
rows. The path in imaginary time must be closed (``a bounce''),
evolving the system  to  the  initial  state~\cite{bounce}.  The
tunneling  rate  is thus given by the sum over all injection and
removal events, $(r_1,t_1)$, $(r_2, t_2)$,
with appropriate phase factors:
  \be
G(w)=
\sum\limits_{1, 2}
\int {\cal D}[u^{(e)}, u^{(h)}]
e^{i\tilde w(r_1-r_2)}e^{ -S^{(e)}_{12} -S^{(h)}_{21}}\ ,
  \ee
where $1$ and $2$ label different positions in the
chain, $\tilde w=2\pi Bw/\Phi_0$, the labels (e) and (h) mark the
electron and hole rows, and $u^{(e,h)}$ is the displacement
field in the rows. It is convenient to rewrite the tunneling
conductivity $G(w)$ as a convolution,
  \be\label{convolution}
G(w)=\int F^{(e)}(\tilde w-s)F^{(h)}(s)ds\ ,
  \ee
where $F^{(e)}$ and $F^{(h)}$ are the {\it structure factors}
of the rows:
  \be\label{structure-factor}
F(w)=
\sum\limits_{1, 2}
\int {\cal D}[u]
e^{i\tilde w(r_1-r_2)}e^{ -S_{12}}\ .
  \ee
The usefulness of using $F(w)$ is that the e- and h-rows may
have different densities or elastic moduli, which will make
spacings and widths of peaks of the conductivity $G(w)$
irregular, while the structure factor of each row still will be
a simple function.
 
Let us calculate the structure factor of one row.
Because of what has been said above we consider only the
long-wavelength displacements $u(x,t)$. Choosing
the units so that $a=1$ and $c=1$, the action for one
row is
  \be
S=\int\!\int \Bigl({\rho\over2}\dot u^2 +{\rho\over2}u^2_x +
\lambda_x \dot u- \dot\lambda u_x \Bigr)\,dx\,dt \ ,
  \ee
where $\rho=m_*/a$, and the Lagrange multiplier $\lambda(x,t)$
is a multivalued function changing by $\pm 2\pi i$ going around
the points $(x_1, t_1)$, $(x_2, t_2)$, respectively. The term
with $\lambda$ is introduced to describe injecting an electron
at $(x_1, t_1)$, and removing it at $(x_2, t_2)$. Qualitatively, the
particles are moving as shown in Fig.~\ref{fig1}.c.
The equation
of motion then is $\nabla^2u=0$, with boundary conditions:\\
  {\it a)\/} The displacement $u(x,t)$ changes by $\pm 1$ going
around the injection and removal points;\\
  {\it b)\/} As a function of time,
$u(x,t)$ is periodic, with period $\beta=1/T$.\\
The solution is given by
  \be
u_0(z)={\rm Im}\ {1\over2\pi}\ln{\sinh \pi T(z-z_1)\over\sinh \pi T(z-z_2)}\ ,
  \ee
where $z=x+it$. Evaluating the action $S[u_0(x,t)]$ one gets
$
{\rho\over 2\pi} \ln
\Bigl|{\beta\over \pi} \sinh
\Bigl({\pi\over\beta}(z_1-z_2)\Bigr)\Bigr|
$, which is equal to
  \be\label{S_0}
S_0=
{\rho\over4\pi}\ln{\beta^2\over\pi^2}
\Bigl(\sinh^2 {\pi
x_{12}\over\beta} +\sin^2 {\pi t_{12}\over\beta} \Bigr)\ ,
  \ee
where $x_{12}=x_1-x_2$, $t_{12}=t_1-t_2$.
 
To average the phase factor in (\ref{structure-factor}), we
write $r_{1,2}$ in terms of the displacement $u(x_{1,2})$
from the mean position $x_{1,2}$, explicitly showing the
particle number in the row:
  \be\label{F(B)}
F(w)=\sum\limits_{x_{12}=n} (-)^n
\langle
e^{i\tilde w(n+u(x_1,t_1)-u(x_2,t_2))}\rangle e^{-S_0}\ ,
  \ee
where the sign is due to exchange of $n+1$ fermions, and
the Debye-Waller factor $\langle...\rangle$ describes
dephasing due to the zero-point and thermal fluctuations.
The gaussian average in (\ref{F(B)})
is performed in a standard fashion,
  \be
\langle...\rangle=e^{i\tilde w n}
e^{-{1\over2}{\tilde w}^2\langle
(u(x_1,-\tau/2)-u(x_2,\tau/2))^2\rangle}
  \ee
where the second exponent equals $-(\tilde w\hbar/\rho a c)^2S_0$.
Taking the extremum of (\ref{S_0}) in the time separation
$t_{12}$ gives optimal $t_{12}=\beta/2$. Finally, restoring $c$
and $a$, the resulting structure factor is
  \be\label{F-result}
F(w)=\sum\limits_n e^{in(\tilde wa+\pi)}
\Bigl({\tilde T\over
\cosh\tilde Tn/2}\Bigr)^{\alpha(\tilde w)}\ ,
  \ee
where $\tilde T=2\pi aT/\hbar c$, $\alpha(\tilde w)= {\rho a^2 c/
2\pi\hbar} +{{\tilde w}^2\hbar/ 2\pi \rho c}$. The structure factor
(\ref{F-result}) has peaks at ${\tilde w}_m=(2\pi/a) m$, in agreement
with (\ref{peaks}). The width of the $m-$th peak scaled in
$a^{-1}$ is given by $\alpha^{1/2}({\tilde w}_m)\,\tilde T$,
which equals
  \be
[ \gamma+m^2/\gamma]^{1/2} \tilde T   \ , \ \ \
\gamma=\rho a^2 c/ 2\pi\hbar\ .
  \ee
The peak width increases with $m$ and is proportional to
temperature.
 
To compare with the experiment\cite{Ray}, we plot the ac
capacitance\\
$C(w)=C/(1+(C\omega/G(w))^2)$ with $G$ given by
(\ref{convolution}), (\ref{F-result}), and $C\omega$ being a
parameter (Fig.~\ref{fig2}). The peaks appear at the transition
between plateaus, and have temperature dependence consistent
with\cite{Ray}.
 
\nin \underline{Qualitative discussion and conclusion}\hskip2mm
The distinction between the Bragg resonances and other
explainations of the observed peaks can be made by studying
thermal shifts of the peaks. In the Bragg condition
(\ref{peaks}) both the strip width $w$ and the crystal period
$a$ depend on $T$, which results in the shifts of the same sign
for all the peaks, and of magnitude proportional to the order of
the peak. The sign of the shifts is determined by the
competition of two factors. The thermal expansion of the crystal
will lead to $a$ increasing, and $w$ decreasing with
temperature. According to (\ref{peaks}), the first effect is
more important, which leads to a shift consistent with
experiment.
 
The effect of disorder is difficult to estimate, but in the
setup\cite{Ray} the sensitivity to disorder should be lower than
in other existing devices: there is a stronger electric field,
and thus a more narrow strip; both sides of the strip are free,
and so the disorder potential will mainly result in the strip
bending with lesser effect on the width.
 
We may conclude by listing several qualitative features of the
experiment correctly described by the above theory. The number
of peaks and their positions in the magnetic field--gate voltage
plane are a result of the Bragg condition. The change in the
height of peaks is a result of the temperature dependence of the
action (\ref{S_0}). The shifts in peak position are explained by
a thermal expansion of the crystal.
 
\acknowledgements
We greatly benefitted from dicussions with Ray Ashoori and Nikolai Zhitenev.

  \begin{figure}
\caption{
  {\it a)\/}
A large collection of magnetic dipoles which repelled each other
was confined between two glass plates, and the glass was tilted;
the in-page gravity component is down\protect\cite{Pieranski}
(courtesy of P.~Pieranski and F.~Rothen).\hskip2mm
  {\it b)\/}
Incompressible strip with few electron and hole rows on the
sides is shown. Beyond the rows are regions of QH fluid. The
coordinate $x$ is along the strip, and $y$ is normal to it.\hskip2mm
  {\it c)\/}
The imaginary time dynamics picture. The world
lines of the tunneling particle and of other particles are
shown.
}
\label{fig1}
  \end{figure}
 
  \begin{figure}
\caption{
The ac capacitance $C(w)=C/(1+(C\omega/G(w))^2)$, with $G$ given
by (\protect\ref{convolution}) and (\protect\ref{F-result}),
plotted as function of the strip width $\tilde w=2\pi wB/\Phi_0$
at several temperatures, offset for clarity. Parameters used:
$C\omega=0.05$, $a^{(e)}=1$, $a^{(h)}=1.68$, $\gamma=2$.
Non-equal e- and h-periods are chosen to demonstrate the effect
of possible period mismatch.
  }
\label{fig2}
  \end{figure}
 
\end{document}